**Woolf et al's "GWAS by subtraction" is not useful for cross-generational Mendelian randomization studies**

David M Evans[1,2,3]*

George Davey Smith[3]

Gunn-Helen Moen[1,2,4,5]

1. Institute for Molecular Bioscience, The University of Queensland, Brisbane, Australia
2. The Frazer Institute, The University of Queensland, 4102, Woolloongabba, QLD, Australia.
3. MRC Integrative Epidemiology Unit, University of Bristol, Bristol, UK
4. Institute of Clinical Medicine, Faculty of Medicine, University of Oslo, Oslo, Norway.
5. Department of Public Health and Nursing, K.G. Jebsen Center for Genetic Epidemiology, NTNU, Norwegian University of Science and Technology, Trondheim, Norway.

*Correspondence to Prof David M Evans, Institute for Molecular Bioscience, University of Queensland, Brisbane, Australia. Email: d.evans1@uq.edu.au.

**Abstract**

Mendelian randomization (MR) is an epidemiological method that can be used to strengthen causal inference regarding the relationship between a modifiable environmental exposure and a medically relevant trait and to estimate the magnitude of this relationship[1]. Recently, there has been considerable interest in using MR to examine potential causal relationships between parental phenotypes and outcomes amongst their offspring. In a recent issue of *BMC Research Notes*, Woolf et al (2023) present a new method, "GWAS by subtraction", to derive genome-wide summary statistics for paternal smoking and other "paternal phenotypes" with the goal that these estimates can then be used in downstream (including two sample) MR studies. Whilst a potentially useful goal, Woolf et al. (2023) focus on the wrong parameter of interest for useful genome-wide association studies and downstream cross-generational MR studies, and the estimator that they derive is neither efficient nor appropriate for such use.

Mendelian randomization (MR) is an epidemiological method that can be used to strengthen causal inference regarding the relationship between a modifiable environmental exposure and a medically relevant trait and to estimate the magnitude of this relationship[1]. Recently, there has been considerable interest in using MR to examine potential causal relationships between parental phenotypes and outcomes amongst their offspring[2-4] (interestingly one of the earliest exemplars of MR was confirmation that antenatal maternal folate was protective against offspring neural tube defects[1]). In a recent issue of *BMC Research Notes*, Woolf et al (2023)[5] present a new method, "GWAS by subtraction", to derive genome-wide summary statistics for paternal smoking and other "paternal phenotypes" with the goal that these estimates can then be used in downstream (including two sample) MR studies[6]. Whilst a potentially useful goal, Woolf et al. (2023) focus on the wrong parameter of interest for useful genome-wide association studies (GWAS) and downstream cross-generational MR studies, and the estimator that they derive is neither efficient nor appropriate for such use.

The paper is peppered with inaccuracies that make it difficult for readers to make sense of what the authors have done. These include (1) mistakes in the equations (e.g. the term "3" is used instead of what is presumably an error term in the first unnumbered equation in the manuscript, and in numbered equation one they seem to conflate the concepts of genetic liability and estimates of genotype-phenotype association); (2) confusing spelling mistakes (e.g. after one occurrence of "inheritance" in the introduction it becomes "inherence" and then it becomes "inertance"; "phenotype" transforms into "phototype", mothers become mouthers and then mathers, etc); (3) confusing descriptions of estimators, simulation procedures and methods (e.g. not indicating estimators with a circumflex; listing the parameters of the binomial distribution "B" in reverse and then including a variable "B" in the simulation equations; simulating the parameter B to be random; the appearance of numerical values such as 0.006 with no explanation, etc). We have refrained from documenting more of these and have restricted ourselves to pointing out some of the issues that are directly relevant to cross-generational genome-wide association studies and the utilization of the SNP-phenotype regression coefficients from these studies in downstream MR analyses. Our letter focuses on the following issues in particular: (1) Woolf et al.'s estimator is not conceptually coherent ; (2) Woolf's estimator is not statistically efficient; (3) Woolf et al.'s formula for the standard error of their estimator is likely to be conservative; (4) Woolf et al.'s empirical analyses use measures of own and maternal smoking that are too different to be combined in their proposed fashion; (5) Woolf et al.'s estimator is unlikely to be useful from the perspective of locus discovery; (6) The empirical MR analyses the authors conduct are equivalent to an (underpowered) MR of own smoking on own lung cancer/emphysema/bronchitis; and (7) the Woolf et al estimator is not a useful quantity for estimating the causal effect of paternal phenotypes on offspring outcomes.

**(1) Woolf et al's estimator of $\beta_{CG\_FP}$ is not conceptually coherent.**

Following Woolf et al., we define the following population level quantities:

$\beta_{CG\_FP}$ = Population regression coefficient of father's phenotype on child genotype (the parameter of interest to Woolf et al)

$\beta_{CG\_MP}$ = Population regression coefficient of mother's phenotype on child genotype

$\beta_{CG\_CP}$ = Population regression coefficient of child's phenotype on child genotype

We assume for the moment that β_CG_FP is a useful quantity to estimate (however, see below on this point). The estimator of β_CG_FP that Woolf et al (2023) propose in their paper is:

$$\hat{\beta}_{CG\_FP} = \hat{\beta}_{CG\_CP} - \hat{\beta}_{CG\_MP}$$

with standard error:

$$\text{se}(\hat{\beta}_{CG\_FP}) = \sqrt{\text{var}(\hat{\beta}_{CG_{CP}}) + \text{var}(\hat{\beta}_{CG\_MP})}$$

where $\hat{\beta}_{CG\_CP}$ is the coefficient from the regression of child phenotype on child genotype in the sample, and $\hat{\beta}_{CG\_MP}$ is the coefficient from the regression of maternal phenotype on child genotype in the sample.

Since Woolf's estimator combines regression coefficients from two generations and collapses across sexes, the estimator implicitly assumes no sex differences in genetic effects or differences in genetic effects across generations. We note that these assumptions are unlikely to hold for a phenotype such as smoking. Cigarette smoking behaviour has changed rapidly in many societies, in some having risen from zero to the majority of the population, and then back to a small minority in less than a century. The distribution of smoking by sex, socioeconomic class, age and ethnicity has similarly fluctuated markedly over this time. This could lead to situations in which it is clear that nonsensical estimates would be made using the GWAS by subtraction approach advocated by the authors. Consider a population in which female smoking is near zero (such as in some African, Asian and Middle Eastern countries currently) and the GWAS by subtraction approach was carried out. The process must, of course, be symmetrical- it should be agnostic to the sex of the parent you have phenotypic data for. If you had paternal smoking data you would produce GWAS by subtraction summary statistics for maternal smoking that could then be used in MR studies to produce estimates of the effects of maternal smoking on offspring and partner outcomes. These would suggest positive influences of maternal smoking on their own and their offspring and partner outcomes even in a situation where no mothers smoked. Indeed, these findings would be similar to the ones that the authors of the paper under discussion have reported with respect to the effects of paternal smoking on their own, their offspring and their partner outcomes.[5,6]

**(2) Woolf et al's estimator of β_CG_FP is not statistically efficient.**

Woolf's estimator is not an efficient estimator of β_CG_FP since it equally weights contributions from the regression of child's phenotype on child's genotype, and mother's phenotype on child's genotype. A superior estimator that is appropriate for the analysis of common genetic variants in large samples uses inverse variance weighting and treats the sample regression coefficient of mother's phenotype on child genotype as one estimate of β_CG_FP, and half the sample regression coefficient of own phenotype on own genotype as another i.e.:

$$\hat{\beta}_{CG\_FP} = \frac{w_1 \times 0.5 \times \hat{\beta}_{CG\_CP} + w_2 \times \hat{\beta}_{CG\_MP}}{w_1 + w_2}$$

where

$$w_1 = \frac{4}{\text{var}(\hat{\beta}_{CG\_CP})}$$

$$w_2 = \frac{1}{\text{var}(\hat{\beta}_{CG\_MP})}$$

with standard error:

$$\text{se}(\hat{\beta}_{CG_{FP}})$$
$$= \sqrt{(\frac{.5w_1}{w_1+w_2})^2 \text{var}(\hat{\beta}_{CG\_CP}) + (\frac{w_2}{w_1+w_2})^2 \text{var}(\hat{\beta}_{CG\_MP}) + \text{cov}(\hat{\beta}_{CG\_CP}, \hat{\beta}_{CG\_MP}) \frac{w_1 w_2}{(w_1+w_2)^2}}$$

Like Woolf et al's estimator, as well as the usual assumptions of no assortative mating, no indirect effects, and no population stratification, this more efficient estimator assumes no difference in genetic effects between males and females, no difference in genetic effects across generations, and that mothers and offspring are measured on the same continuous phenotype on the same measurement scale (or that their measurements can be transformed to the same continuous scale).

**(3) Woolf et al's formula for the standard error of their estimator is only correct under independence and therefore likely to be conservative in their own and other real-world applications.** The formula for the standard error that Woolf et al uses is only correct under the assumption that $\hat{\beta}_{CG\_CP}$ and $\hat{\beta}_{CG\_MP}$ are independent (i.e. estimates of SNP-phenotype association are derived from completely independent samples and/or there exists no residual covariance between maternal phenotype and offspring phenotype over and above the SNP-phenotype associations). Rather, the correct formula for the standard error of their estimator allowing for covariance between the regression coefficients is:

$$\text{se}(\hat{\beta}_{CG\_FP}) = \sqrt{\text{var}(\hat{\beta}_{CG\_CP}) + \text{var}(\hat{\beta}_{CG\_MP}) - 2\text{cov}(\hat{\beta}_{CG\_CP}, \hat{\beta}_{CG\_MP})}$$

We note that, if necessary, the covariance term in the above formula can be estimated by e.g. bivariate LD score regression[7]:

$$\widehat{\text{cov}}(\hat{\beta}_{CG\_CP}, \hat{\beta}_{CG\_MP}) \approx \frac{N_S}{\sqrt{N_{CG\_CP} N_{CG\_MP}}} \rho \sqrt{\text{var}(\hat{\beta}_{CG\_CP}) \text{var}(\hat{\beta}_{CG\_MP})}$$
$$= \widehat{int} \times \text{se}(\hat{\beta}_{CG\_CP}) \text{se}(\hat{\beta}_{CG\_MP})$$

where $N_{CG\_CP}$ is the number of individuals in the regression of child phenotype on child genotype, $N_{CG\_MP}$ is the number of mothers in the regression of maternal phenotype on child genotype, $N_S$ is the effective sample overlap between the two regressions, $\rho$ is the correlation between maternal and child phenotypes, and $\widehat{int}$ is the bivariate LD score regression intercept[8]. Because the residual covariance between maternal and offspring phenotypes is likely to be positive, any sample overlap should result in a smaller standard error than what Woolf et al proposes. The corollary is, other important considerations aside, that #2 and #3 imply that Woolf et al's empirical analyses of smoking in the UK Biobank are likely to be conservative.

**(4) Woolf et al's empirical analyses use measures of smoking that are different.** Woolf et al's estimator assumes that offspring and maternal phenotypes have been measured on the same (continuous linear) scale, however, their empirical analyses in the UK Biobank use phenotypes where this assumption is highly questionable (i.e. a GWAS of "lifetime smoking index" versus a maternal GWAS of smoking/non-smoking). It is not appropriate to combine the regression coefficients of different phenotypes in the fashion that Woolf et al suggest regardless of whether the phenotypes (or the regression coefficients) have been standardized.

**(5) Estimation of $\hat{\beta}_{CG\_FP}$ is unlikely to be useful from the perspective of genetic locus discovery.**

As shown above, Woolf et al's estimator is not efficient. Since there is an implicit assumption of no generational effects, far more useful would be an estimate of own genotype-own phenotype association that combines information from the GWAS of own phenotype with information from a GWAS of maternal phenotype. We and others have shown various ways of incorporating parental and offspring GWAS to improve the power of locus discovery[9-12].

**(6) The Empirical MR analysis the authors conduct is equivalent to an (underpowered) MR of own smoking on own lung cancer/emphysema/bronchitis.**

The authors use $\hat{\beta}_{CG\_FP}$ in downstream MR analyses to investigate whether paternal smoking causes paternal lung cancer/bronchitis/emphysema. However, assuming the absence of generational effects, this analysis is basically the same (albeit a less powerful way) of investigating whether (own) smoking causes (own) lung cancer.

**(7) $\hat{\beta}_{CG\_FP}$ is not a useful quantity for estimating the causal effect of paternal phenotypes on offspring outcomes.**

A far more interesting empirical question than the one addressed in Woolf et al involves estimating the causal effect of paternal smoking on offspring phenotypes. Indeed the authors have gone on to do precisely this in a subsequent publication using their summary statistics[6]. Unfortunately, the summary statistics derived from Woolf et al cannot be used for this purpose. Rather, for these sorts of analyses, what is required is an estimate of the association between paternal genotype and offspring outcome (conditional on offspring genotype). These estimates could be obtained by e.g. conditional analysis of genotyped father-offspring pairs/parent-offspring trios, and then used in one or two sample MR studies. Indeed, similar investigations have been proposed (and performed) in MR studies of maternal exposures and offspring outcomes[2,4,13]. In other words, "conditional GWAS" rather than "GWAS by subtraction" is required for valid cross-generational MR studies.

Alternatively, an MR by proxy design could be used to investigate the causal effects of parental exposures on offspring outcomes, in which offspring genotype proxies for maternal (or paternal) genotype[14]. However, in order for this sort of design to be informative for causality, it requires data on the parental phenotype and an informative genotype by environment interaction. Indeed, even if these prerequisites are satisfied, there may still be limitations with respect to which parental exposure-offspring phenotypes can be validly examined for causality. For example, the first study to employ this design[14], used a SNP in the offspring *CHRNA5* gene (i.e. a genetic variant related to smoking) to proxy maternal genotype at the same locus, and investigate the relationship between maternal smoking during pregnancy and offspring birthweight. The authors reasoned that any association between offspring SNP and offspring birthweight due to a causal effect of maternal smoking on offspring birthweight, would only be observed in offspring whose mothers smoked during pregnancy. Indeed, it was critical that maternal smoking status during pregnancy was available in order for the authors' design to yield valid causal inferences. However, the same design would have less validity to examine the effect of maternal smoking during pregnancy and e.g. later life outcomes as such an association could be generated through post-natal maternal smoking and/or offspring smoking[15].

Finally, we note that the authors have failed to mention several recent methodological extensions of genetic association studies that permit estimation of maternal, paternal and offspring genetic effects when parent-child duos/trios are not readily available, and that can be used in downstream cross-generational MR studies[2,8,16-21]. These includes approaches that statistically impute parental genotypes using related individuals' genotypes in situations where parents have not been physically genotyped[19-21]. There is a century old literature from animal and human genetics that discusses the value of relatives with genotypes but no phenotype, or with phenotypes but no genotypes[22], but the notion of strengthening causal inference through simply thinking of a parental group - with neither phenotype nor genotype data available for them (whether imputed or otherwise) – is alchemic. Extensive discussion of cross-generational MR methodology as well as some recent developments are available elsewhere[2,3].

**References**

1. Smith, G.D. & Ebrahim, S. 'Mendelian randomization': can genetic epidemiology contribute to understanding environmental determinants of disease? *Int J Epidemiol* **32**, 1-22 (2003).
2. Evans, D.M., Moen, G.H., Hwang, L.D., Lawlor, D.A. & Warrington, N.M. Elucidating the role of maternal environmental exposures on offspring health and disease using two-sample Mendelian randomization. *Int J Epidemiol* **48**, 861-875 (2019).
3. Lawlor, D. *et al.* Using Mendelian randomization to determine causal effects of maternal pregnancy (intrauterine) exposures on offspring outcomes: Sources of bias and methods for assessing them. *Wellcome Open Res* **2**, 11 (2017).
4. Moen, G.H. *et al.* Mendelian randomization study of maternal influences on birthweight and future cardiometabolic risk in the HUNT cohort. *Nat Commun* **11**, 5404 (2020).
5. Woolf, B., Sallis, H.M., Munafo, M.R. & Gill, D. Deriving GWAS summary estimates for paternal smoking in UK biobank: a GWAS by subtraction. *BMC Res Notes* **16**, 159 (2023).
6. Woolf, B., Rajasundaram, S., Gill, D., Sallis, H.M. & Munafo, M.R. Assessing the causal effects of environmental tobacco smoke exposure: A meta-analytic Mendelian randomization study. *medRxiv*, https://doi.org/10.1101/2023.03.30.23287949 (2023).
7. Bulik-Sullivan, B. *et al.* An atlas of genetic correlations across human diseases and traits. *Nat Genet* **47**, 1236-41 (2015).
8. Wu, Y. *et al.* Estimating genetic nurture with summary statistics of multigenerational genome-wide association studies. *Proc Natl Acad Sci U S A* **118**(2021).
9. de la Fuente, J., Grotzinger, A.D., Marioni, R.E., Nivard, M.G. & Tucker-Drob, E.M. Integrated analysis of direct and proxy genome wide association studies highlights polygenicity of Alzheimer's disease outside of the APOE region. *PLoS Genet* **18**, e1010208 (2022).
10. Hujoel, M.L.A., Gazal, S., Loh, P.R., Patterson, N. & Price, A.L. Liability threshold modeling of case-control status and family history of disease increases association power. *Nat Genet* **52**, 541-547 (2020).
11. Liu, J.Z., Erlich, Y. & Pickrell, J.K. Case-control association mapping by proxy using family history of disease. *Nat Genet* **49**, 325-331 (2017).
12. Hwang, L.D. *et al.* Direct and INdirect effects analysis of Genetic lOci (DINGO): A software package to increase the power of locus discovery in GWAS meta-analyses of perinatal phenotypes and offspring traits influenced by indirect genetic effects. *medRxiv*, https://www.medrxiv.org/content/10.1101/2023.08.22.23294446v1 (2023).
13. Wang, G. *et al.* Investigating a Potential Causal Relationship Between Maternal Blood Pressure During Pregnancy and Future Offspring Cardiometabolic Health. *Hypertension* **79**, 170-177 (2022).
14. Yang, Q., Millard, L.A.C. & Davey Smith, G. Proxy gene-by-environment Mendelian randomization study confirms a causal effect of maternal smoking on offspring birthweight, but little evidence of long-term influences on offspring health. *Int J Epidemiol* **49**, 1207-1218 (2020).
15. Hwang, L.D. & Evans, D.M. Commentary: Proxy gene-by-environment Mendelian randomization for assessing causal effects of maternal exposures on offspring outcomes. *Int J Epidemiol* **49**, 1218-1220 (2020).

16. Warrington, N.M. *et al.* Maternal and fetal genetic effects on birth weight and their relevance to cardio-metabolic risk factors. *Nat Genet* **51**, 804-814 (2019).
17. Warrington, N.M., Freathy, R.M., Neale, M.C. & Evans, D.M. Using structural equation modelling to jointly estimate maternal and fetal effects on birthweight in the UK Biobank. *Int J Epidemiol* **47**, 1229-1241 (2018).
18. Warrington, N.M., Hwang, L.D., Nivard, M.G. & Evans, D.M. Estimating direct and indirect genetic effects on offspring phenotypes using genome-wide summary results data. *Nat Commun* **12**, 5420 (2021).
19. Hwang, L.D. *et al.* Estimating indirect parental genetic effects on offspring phenotypes using virtual parental genotypes derived from sibling and half sibling pairs. *PLoS Genet* **16**, e1009154 (2020).
20. Tubbs, J.D., Hwang, L.D., Luong, J., Evans, D.M. & Sham, P.C. Modeling Parent-Specific Genetic Nurture in Families with Missing Parental Genotypes: Application to Birthweight and BMI. *Behav Genet* **51**, 289-300 (2021).
21. Young, A.I. *et al.* Mendelian imputation of parental genotypes improves estimates of direct genetic effects. *Nat Genet* **54**, 897-905 (2022).
22. Visscher, P.M. & Duffy, D.L. The value of relatives with phenotypes but missing genotypes in association studies for quantitative traits. *Genet Epidemiol* **30**, 30-6 (2006).